\documentclass[12pt]{article}
\usepackage{graphicx}
\usepackage{amssymb}
\usepackage{amsmath}

\begin{document}

\begin{center}
{\Large{\bf Kinetics of Surface Enrichment: A Molecular Dynamics Study}} \\
\ \\
\ \\
by \\
Prabhat K. Jaiswal$^1$, Sanjay Puri$^1$, and Subir K. Das$^{2}$ \\
$^1$School of Physical Sciences, Jawaharlal Nehru University, New Delhi -- 110067, India. \\
$^2$Theoretical Sciences Unit, Jawaharlal Nehru Centre for Advanced Scientific Research, 
Jakkur, Bangalore -- 560064, India.
\end{center}

\begin{abstract}
We use molecular dynamics (MD) to study the kinetics 
of surface enrichment (SE) in a stable homogeneous mixture ($AB$), 
placed in contact with a surface which preferentially attracts $A$.
The SE profiles show a characteristic double-exponential behavior 
with two length scales: $\xi_-$, which rapidly saturates to its 
equilibrium value, and $\xi_+$, which diverges as a power-law 
with time ($\xi_+ \sim t^\theta$). We find that hydrodynamic 
effects result in a crossover of the growth exponent from 
$\theta\simeq 0.5$ to $\theta\simeq1.0$. There is also a 
corresponding crossover in the growth dynamics of the SE-layer 
thickness.
\end{abstract}
                             
\section{Introduction}
\label{intro}
There is a wide range of physical phenomena associated with binary 
($AB$) mixtures in contact with a surface ($S$). Typically, the surface 
has a preferential attraction for one of the components of the mixture 
(say, $A$). A rich phase diagram arises, depending on the interactions 
between $AB, AS, BS$ and the temperature $T$ 
\cite{cahn77,mef84,mef86,gennes85,sd88}. For simplicity, we focus on 
a semi-infinite geometry. If the bulk system is below the miscibility 
gap, the $AB$ interface meets the surface at a contact angle $\theta$, 
which is determined by Young's equation \cite{young}:
\begin{equation}
 \gamma_{AB}\,\mathrm{cos}\theta = \gamma_{BS} - \gamma_{AS},
 \label{eq1:young}
\end{equation}
where $\gamma_{\alpha\beta}$ denotes the surface tension between $\alpha$ 
and $\beta$. For $\gamma_{AB} > \gamma_{BS}-\gamma_{AS}$, the surface 
undergoes \textit{partial wetting}. For $\gamma_{AB} < \gamma_{BS} - \gamma_{AS}$, 
Young's equation does not have a solution and the surface is 
\textit{completely wet} in equilibrium, i.e., the $B$-rich 
phase is expelled from the surface. On the other hand, if the bulk system 
is above the miscibility gap, there will be no macroscopic phase 
separation. Nevertheless, the surface is enriched in the preferred 
component -- the degree of enrichment depends upon various interaction 
strengths. Both the problems of \textit{wetting} and 
\textit{surface enrichment} are of great scientific and technological 
importance. 

We have a long-standing interest in the kinetics of mixtures 
at surfaces. Again, a rich set of problems arises in this context. 
Consider a homogeneous $AB$ mixture (at high temperatures, $T=\infty$) 
placed in contact with a flat surface (located at $z=0$) which prefers 
$A$. If the system is suddenly quenched below the 
miscibility gap at time $t=0$, it undergoes phase separation in the bulk 
\cite{bray94,pw09}, and segregates into $A$-rich and $B$-rich domains. 
Simultaneously, the surface is wetted by $A$. The interplay of 
these two kinetic processes, i.e., phase separation and wetting, results 
in surface-directed spinodal decomposition (SDSD) waves which originate 
at the surface and propagate into the bulk. There has been intense 
experimental \cite{jnkbw91,kdkk93,krausch95,gk03} and theoretical 
\cite{pf97,puri05,bpdh10} interest in SDSD. Alternatively, we can consider 
quenches to temperatures above the miscibility gap or to the metastable 
region of the phase diagram \cite{pb01}. In this case, the 
mixture is stable and continues to be homogeneous for large $z$. However, 
the surface becomes enriched in the preferred component, resulting in a 
time-dependent surface-enrichment (SE) profile, which propagates into the bulk. 
This phenomenon of SE is well known in the context of 
polymer blends \cite{yl06,jkrss89,jk90,gnp97,gc97,wds03} and biopolymer 
mixtures \cite{mouritsen}, binary alloys \cite{blakely79,lkv10}, and 
the wetting of surfaces of fluids \cite{gennes85,gbr90}.

In this paper, we undertake a comprehensive molecular dynamics (MD) 
study of the kinetics of SE. This paper has two primary goals. 
First, we would like to test the 
theoretical results for enrichment kinetics (summarized later) via a 
realistic microscopic MD simulation. The available numerical studies 
of this problem have focused on Langevin simulations of phenomenological 
models. Second, we would like to examine whether hydrodynamic effects 
(which are naturally incorporated in our MD simulations) have any 
effect on SE kinetics. After all, many experiments in this area 
involve fluid mixtures or polymer mixtures, where velocity fields 
play a significant role. In the context of phase-separation kinetics, 
we know that hydrodynamic effects drastically alter the late-stage 
dynamics -- both without surfaces \cite{pw09,bray94,kcpdb01,wc01,adp10} 
and with surfaces \cite{tanaka01,bpl01,jpd}.

This paper is organized as follows. Section~\ref{sec:detail-md} 
describes the details of our MD simulations. In Sec.~\ref{sec:results}, 
we present comprehensive numerical results and compare them with 
theoretical predictions. Finally, Sec.~\ref{sec:sum} concludes this 
paper with a summary and discussion.

\section{Details of Molecular Dynamics Simulations}
\label{sec:detail-md}
Following Das et al. \cite{dphb06}, we consider a binary ($AB$) fluid mixture of 
point particles confined in a 
rectangular box of volume $V=L\times L\times D$. We apply periodic 
boundary conditions in the $x$ and $y$ directions, while impenetrable 
walls or surfaces are present at $z=0$ and $z=D$. These surfaces 
give rise to an integrated Lennard-Jones (LJ) potential ($\alpha=A,B$):
\begin{equation}
u_w(z)=\frac{2\pi n\sigma^3}{3} \left[\frac{2\epsilon_r}{15}{\left(\frac{\sigma}{z^\prime}\right)}^9
-\delta_\alpha\epsilon_a{\left(\frac{\sigma}{z^\prime}\right)}^3\right],
\label{eq2:intlj}
\end{equation}
where $n$ is the reference density of the fluid,
and $\sigma$ is the LJ diameter of the particles. In Eq.~(\ref{eq2:intlj}), 
$\epsilon_r$ and $\epsilon_a$ are the strengths of the 
repulsive and attractive parts of the wall potentials. For the wall 
at $z=0$, we choose $\delta_A=1$ and $\delta_B=0$, so there is only a 
repulsion for $B$ particles, while $A$ particles are attracted 
at large distances and repelled at short distances. 
For the wall at $z=D$, we set $\delta_A=\delta_B=0$, 
so that both $A$ and $B$ particles feel only repulsion. Further, 
$z^\prime=z+\sigma/2$ for the wall at $z=0$, and 
$z^\prime=D+\sigma/2-z$ for the wall at $z=D$. Therefore, the 
singularities of $u_w(z)$ do not occur within the box range 
$0\leq z\leq D$, but rather at $z= -\sigma/2$ and 
$z=D+\sigma/2$, respectively. 

The particles in the system interact with LJ potentials:
\begin{equation}
 u(r_{ij})=4\epsilon_{\alpha\beta}\left[\left(\frac{\sigma}
{r_{ij}}\right)^{12}-\left(\frac{\sigma}{r_{ij}}\right)^{6}\right],
\label{eq3:lj}
\end{equation}
where $r_{ij}=|\vec{r}_i-\vec{r}_j|$; $\alpha,\beta=A,B$. The energy 
scales are
\begin{equation}
 \epsilon_{AA}=\epsilon_{BB}=2\epsilon_{AB}=\epsilon,
 \label{eq4:ljpara}
\end{equation}
for which the equilibrium phase behavior is well studied 
\cite{dhb03,dhbfs06,dfshb06}.
We use the truncated, shifted and force-corrected LJ potential with 
$r_{ij}=r_c=2.5 \sigma$ \cite{at87}. We consider the case with 
equal numbers of $A$ and $B$ particles ($N_A=N_B=N/2$), and their masses 
are set to be equal, $m_A=m_B=m=1$. We also set $\sigma=1$, $\epsilon=1$ 
and $k_B=1$, such that the MD time unit is
\begin{equation}
 t_0={\left(\frac{m\sigma^2}{48\epsilon}\right)}^{1/2}=\frac{1}{\sqrt{48}}.
 \label{eq5:mdtime}
\end{equation}
We work with a high-density liquid having $n=N/V=1$, which makes the system 
incompressible. Notice that crystallization is not a problem 
at the temperatures of interest. Finally, simulations were run for 
three surface energy strengths -- $\epsilon_a=1.2, 2.0, 3.0$ with 
$\epsilon_r=0.5$.

For our study, we chose $L=32$ and $D=64$ ($N=66536$ particles). 
As the bulk remains homogeneous, the lateral size $L$ (in the 
$x, y$ directions) is not severely constrained. However, in the 
direction perpendicular to the surface located at $z=0$, we 
need sufficiently large $D$ to ensure decay of the enrichment 
profiles as $z \rightarrow D$. For the range of
times studied here ($t \leq 7000$), test runs with other linear 
dimensions showed that $D=64$ is large enough to
eliminate finite-size effects, within the limits of our statistical
accuracy. For a smaller system size ($D=32$),
we encounter finite-size effects by $t \simeq 3000$.
The statistical quantities presented here were obtained 
as averages over 50 independent runs. 
The MD runs were carried out using the standard Verlet velocity 
algorithm \cite{bc96}, with a time-step $\Delta t=0.07$ in 
MD units given by Eq.~(\ref{eq5:mdtime}), and the Nos\'{e}-Hoover 
thermostat \cite{bc96}. We prepared the homogeneous initial state 
for a run by equilibrating the mixture of $N$ particles at high $T$, 
in the specified volume with periodic boundary conditions in all directions. 
At time $t=0$, the system is quenched to 
$T=2.0 \simeq 1.41 T_c~(T_c \simeq 1.423)$ \cite{dhbfs06,dfshb06}, and 
surfaces are introduced at $z=0,D$.

\section{Theoretical Background and Numerical Results}
\label{sec:results}
We are interested in the time-dependent 
morphology which arises during SE. We 
characterize the morphology via laterally-averaged depth profiles 
and their various properties, e.g., surface value of order parameter, 
zero-crossings, moments, etc. Before proceeding, it is useful to 
summarize theoretical results in this context.
\subsection{Theoretical Background}
\label{theory}
Jones et al. 
\cite{jkrss89, jk90} have studied 
the kinetics of SE for polymer mixtures, and found 
that the enrichment profiles are characterized by diffusive 
length scales. Numerical studies of this problem (with both short-ranged 
and long-ranged surface fields) have been conducted by Jiang and 
Ebner \cite{je89} using MC simulations, and Toral and Chakrabarti 
\cite{tc91} via Langevin simulations. Binder and Frisch 
\cite{bf91} studied SE in a linearized Ginzburg-Landau (GL) model with 
a delta-function (short-ranged) surface potential. Their GL model was derived 
by coarse-graining the Kawasaki spin-exchange kinetic Ising model in 
the presence of a surface. Their results are expected to be valid 
in the weak-field regime, where 
the concentration variations are sufficiently weak 
that nonlinear effects can be neglected. Puri and Frisch (PF) 
\cite{pf93} have also studied the case with a short-ranged surface 
potential in the framework of a linear theory. They also 
undertook Langevin simulations of the corresponding GL model for both 
weak and strong surface fields. In the latter case, linear theory 
is no longer applicable as the order parameter value at the surface 
becomes appreciably large. PF demonstrated that the morphological 
features predicted by linear theory are also seen in the nonlinear 
regime. These results were extended to the case with an arbitrary 
surface potential $V(z)$ by Frisch et al. (FPN) \cite{fpn99}. Here, 
we briefly summarize the results of PF and FPN.

Let us first discuss the modeling of segregation kinetics 
at surfaces \cite{puri05,bpdh10}.
The bulk order parameter satisfies the Cahn-Hilliard (CH) equation:
\begin{eqnarray}
 \frac{\partial}{\partial t}\phi(\vec{\rho},z,t)=-\nabla^2\Big[ 
 \mathrm{sgn}(T_c-T)\phi-\phi^3
 +\frac{1}{2}\nabla^2\phi-V(z)\Big],\quad z>0. 
 \label{eq6:bulk}
\end{eqnarray}
Here, all the quantities have been rescaled into dimensionless units
\cite{model}. The order parameter 
$\phi(\vec{r},t) \simeq n_A(\vec{r},t)-n_B(\vec{r},t)$, where 
$n_\alpha(\vec{r},t)$ denotes the density of species $\alpha$ at 
space-point $\vec{r}$ and time $t$. We have decomposed coordinates as 
$\vec{r}\equiv (\vec{\rho},z)$, where $\vec{\rho}$ and $z$ denote 
coordinates parallel and perpendicular to the surface located 
at $z=0$. The function $\textrm{sgn}(x)=1$ for $x>0$ and $-1$ 
for $x<0$. The surface potential $V(z)\,(<0)$ is chosen so that it 
enriches the surface in $A$.

Equation~(\ref{eq6:bulk}) must be supplemented by two boundary 
conditions at $z=0$, as it is a fourth-order partial differential 
equation. Since the surface value of the order parameter is not a 
conserved quantity, we assume a nonconserved relaxational kinetics 
with time-scale $\tau_0$ for this quantity at the surface:
\begin{equation}
 \tau_0 \frac{\partial}{\partial t}\phi(\vec{\rho},0,t)=h_1+g\phi(\vec{\rho},0,t)
+\gamma \frac{\partial}{\partial z}\phi(\vec{\rho},z,t)\bigg|_{z=0}
+\tilde{\gamma}\nabla_\|^2\phi(\vec{\rho},0,t),
\label{eq7:bc1}
\end{equation}
where $h_1=-V(0)$, and $g,\gamma,\tilde{\gamma}$ are phenomenological parameters, 
which are related to the bulk correlation length \cite{model}. 
Finally, we implement the no-flux boundary condition at 
the surface, which enforces order parameter conservation:
\begin{eqnarray}
 0=\dfrac{\partial}{\partial z}\left[ 
 \mathrm{sgn}(T_c-T)\phi-\phi^3 
 +\frac{1}{2}\nabla^2\phi-V(z)\right]\bigg|_{z=0}. 
 \label{eq8:bc2}
\end{eqnarray}
Equations~(\ref{eq6:bulk})-(\ref{eq8:bc2}) describe both SDSD (when 
$T<T_c$) and SE (when $T>T_c$), as long as the dynamics 
is diffusive. This is appropriate for phase separation in solid 
mixtures or the early stages of segregation in polymer mixtures. 
However, most experiments involve fluid mixtures, where hydrodynamic 
effects play an important role in the intermediate and late stages 
of phase separation. At a phenomenological level, hydrodynamic effects 
can be incorporated via the Navier-Stokes equation for the velocity 
field -- the resultant coupled equations are known as {\it Model H} 
\cite{tanaka01,hh77}. This must be supplemented by appropriate boundary 
conditions at the surfaces. Alternatively, one can consider 
microscopic models of fluid mixtures at a surface, which naturally 
incorporate the fluid velocity field. We adopt the latter strategy 
in this paper, and use MD simulations to study SE in fluid mixtures.

For reference, it is useful to summarize results for the diffusive case. 
We consider the model in Eqs.~(\ref{eq6:bulk})-(\ref{eq8:bc2}) 
for $T>T_c$, as we are interested in the kinetics of SE. 
In this case, the order parameter field remains 
homogeneous in the direction parallel to the surface, i.e., we can 
neglect the $\vec{\rho}\,$-dependence of the order parameter 
$\phi(\vec{\rho},z,t)\simeq \phi(z,t)$. FPN have solved the linearized 
version of this model, which is appropriate when $\phi$ is small and 
the $\phi^3$-term can be neglected in Eqs.~(\ref{eq6:bulk}) and 
(\ref{eq8:bc2}). Seeing that the bulk remains homogeneous, there is 
a significant enhancement of $\phi$ only for $z\simeq 0$. Thus, the 
linearized model is valid for \textit{weak} surface fields. In this 
case, the results of FPN are as follows.

The SE profiles have a double-exponential form 
\cite{bf91}:
\begin{eqnarray}
 \phi(z,t)\simeq B_-(t)\,\mathrm{e}^{-z/\xi_-(t)}-B_+(t)\,\mathrm{e}^{-z/\xi_+(t)}, 
 \label{eq9:expfit}
\end{eqnarray}
where the amplitudes $B_-(t), B_+(t)>0$. Notice that the conservation
constraint dictates that $B_-\xi_-=B_+\xi_+$. The quantities $B_-(t)$ 
and $\xi_-(t)$ rapidly saturate to their equilibrium values which 
depend on the surface potential:
\begin{eqnarray}
 B_-(t) &\simeq& a_1, \nonumber \\
 \xi_-(t) &\simeq& b_1.
 \label{eq10:minus}
\end{eqnarray}
The other length scale $\xi_+(t)$ grows diffusively with time, and 
$B_+(t)$ shows a corresponding decay:
\begin{eqnarray}
 B_+(t) &\simeq& a_2\,t^{-1/2}, \nonumber \\
 \xi_+(t) &\simeq& b_2\,t^{1/2}.
 \label{eq11:plus}
\end{eqnarray}

The other properties of the enrichment profiles are obtained 
from Eq.~(\ref{eq9:expfit}). An experimentally important quantity is 
the time-dependence of $\phi(z=0,t)$, the value of the order 
parameter at the surface \cite{gnp97,gc97,wds03}:
\begin{equation}
 \phi(0,t)\simeq a_1 - a_2\,t^{-1/2}.
 \label{eq12:phizero}
\end{equation}
Thus, $\phi(0,t)$ saturates diffusively to its equilibrium value. 
The thickness of the enrichment layer is measured as the first zero 
of the double-exponential profile. This increases logarithmically 
with time:
\begin{eqnarray}
 Z_0(t) &\simeq& \frac{\xi_+\xi_-}{\xi_+ - \xi_-}
\ln\left(\frac{B_-}{B_+}\right) \nonumber \\
&\simeq&\frac{\xi_-}{2}\ln\left(\frac{t}{\tau}\right),\quad\tau={\left(\frac{a_2}{a_1}\right)}^2.
\label{eq13:1stzero}
\end{eqnarray}
Finally, consider the time-dependence of the profile moments:
\begin{equation}
 \langle z^m \rangle = \int_0^\infty \mathrm{d}z\,z^m \phi(z,t)\simeq
\frac{1}{m!}\left(B_-\xi_-^{m+1}-B_+\xi_+^{m+1}\right).
 \label{eq14:moments}
\end{equation}
Therefore, the asymptotic ($t\rightarrow\infty$) behavior of $\langle z^m\rangle$ is
$\langle z^m \rangle \sim t^{m/2}$.

These results apply for diffusive transport and are universal
for a wide set of potentials. They have been obtained 
in the context of a linear theory, and there is a range of 
\textit{weak} surface fields where the analytical results obtained from 
the linear model agree well with the numerical solution of the nonlinear 
Eqs.~(\ref{eq6:bulk})-(\ref{eq8:bc2}) \cite{fpn99}. As the field strength 
is increased, the validity of linear theory breaks down in the vicinity of 
the surface as the degree of enrichment becomes larger. Nevertheless, 
FPN demonstrated that the diffusive behavior of various profile 
characteristics is unaffected, even in the strongly nonlinear regime. 

As stated earlier, we have undertaken MD simulations to 
examine whether hydrodynamic effects have any impact on the 
above phenomenology. Let us next present results from these simulations. 

\subsection{Numerical Results}
\label{results}
\begin{figure}[!htbp]
\begin{center}
\includegraphics*[width=0.75\textwidth]{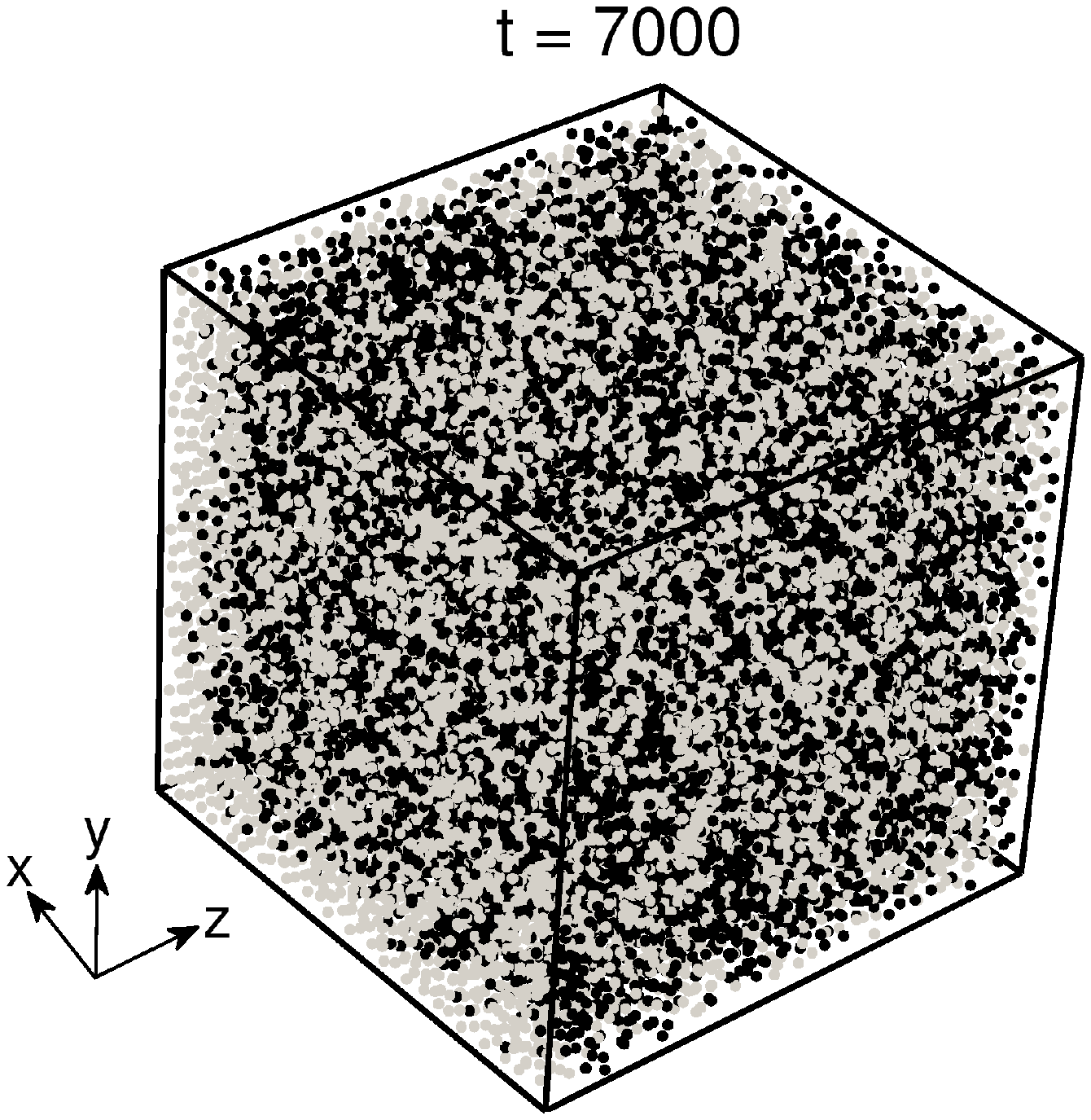}
\end{center}
\caption{Snapshot of surface enrichment (SE) in a binary ($AB$) 
Lennard-Jones (LJ) mixture, which is confined in a box of size 
$L\times L\times D$ with $L=32$, $D=64$. (All lengths are measured 
in units of the LJ diameter.) Periodic boundary conditions are applied 
in the $x,y$ directions, while an impenetrable $L\times L$ surface at 
$z=0$ attracts the $A$-particles. For clarity, we show only part of the
simulation box with $z \in [0,32]$. The initial condition for this run 
consisted of a random mixture of equal amounts of $A$ and $B$ particles 
($N_A=N_B=32768$). Time is measured in dimensionless LJ units. For 
further details of the MD simulation, see Sec.~\ref{sec:detail-md}. 
The $A$-particles are marked gray, and the $B$-particles are marked black. 
The snapshot corresponds to $t=7000$ for the surface potential in 
Eq.~(\ref{eq2:intlj}) with $\epsilon_a=3.0, \epsilon_r=0.5$. The $A$-rich 
enrichment layer is formed at $z=0$, while for large $z$, 
the stable mixture continues to be homogeneous.}
\label{fig:fig1}
\end{figure}

\begin{figure}[!htbp]
\begin{center}
\includegraphics*[width=0.95\textwidth]{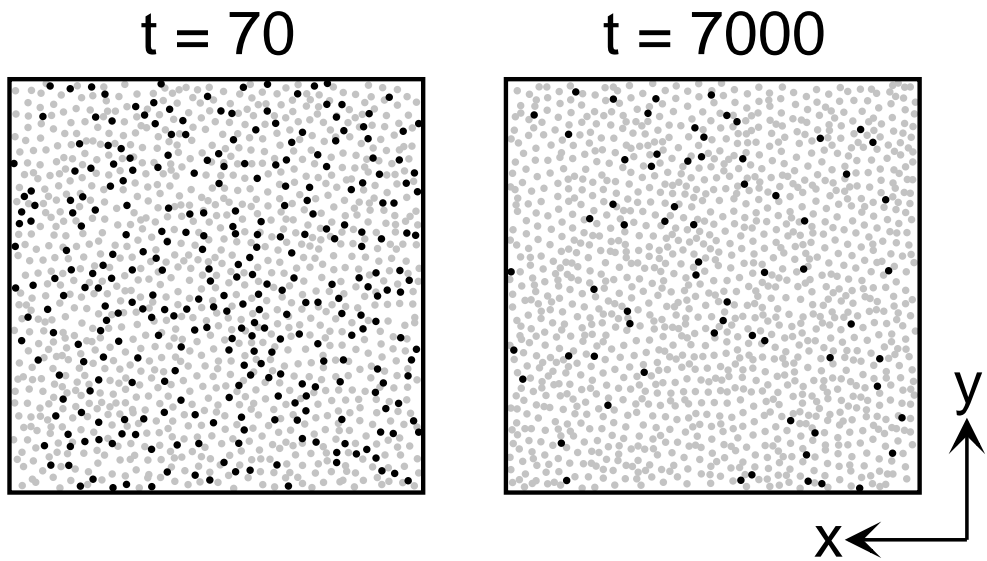}
\end{center}
\caption{Cross-section at $z=0$ of the SE snapshots at time $t=70$ and 
$7000$. The $A$-particles are marked gray and the $B$-particles are 
marked black. Other simulation details are the same as for 
Fig.~\ref{fig:fig1}.}
\label{fig:fig2}
\end{figure}

Figure~{\ref{fig:fig1}} shows a three-dimensional snapshot of SE in 
a binary ($AB$) mixture at $t=7000$. The surface 
field strengths are $\epsilon_a=3.0$ and $\epsilon_r=0.5$ in Eq.~(\ref{eq2:intlj}). 
We see the formation of an $A$-rich (marked in gray) layer at the surface 
($z=0$), resulting in a time-dependent SE profile which 
propagates into the bulk. However, in the bulk (large $z$), the 
thermodynamically stable mixture continues to be homogeneous. 
In Fig.~{\ref{fig:fig2}}, we show cross-sections of the 
snapshots at $z=0$ for $t=70,7000$. The cross-section shows all $A$ atoms 
(marked gray) and all $B$ atoms (marked black) lying in the interval 
$z\in[0,\sigma]$.

\begin{figure}[!htbp]
\begin{center}
\includegraphics*[width=0.75\textwidth]{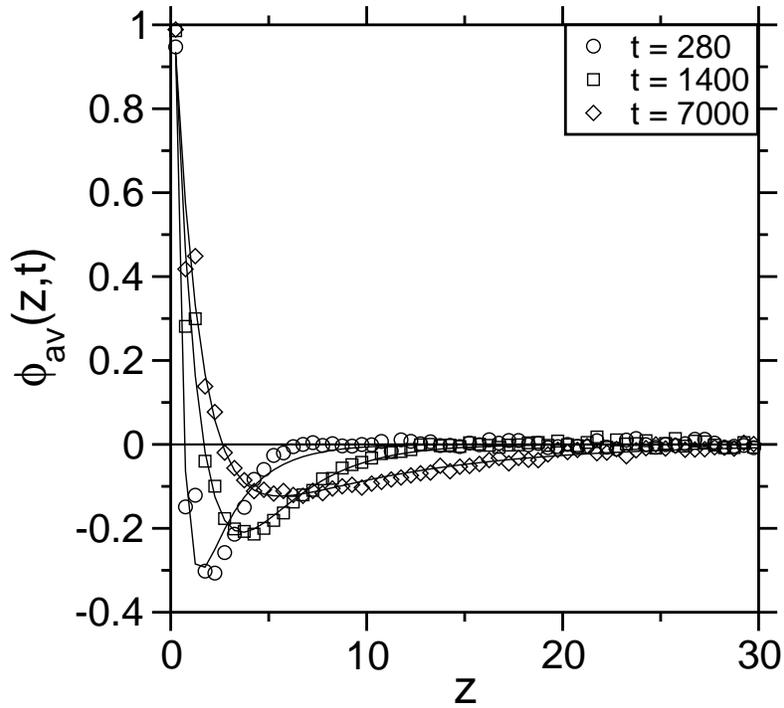}
\end{center}
\caption{Laterally-averaged order parameter profiles 
[$\phi_{\textrm{av}}(z,t)$ vs. $z$] at the dimensionless times $t=280,1400,7000$. 
The surface potential is given by Eq.~(\ref{eq2:intlj}) with 
$\epsilon_a=2.0,\,\epsilon_r=0.5$. The double-exponential fits for the 
SE profiles are shown as solid lines.}
\label{fig:fig3}
\end{figure}

In Fig.~{\ref{fig:fig3}}, we show the temporal evolution of the 
laterally-averaged order parameter profiles $\phi_{\mathrm{av}}(z,t)$ vs. $z$, 
obtained from our MD simulations \cite{model}. The order parameter 
is defined in terms
of the local densities as $\phi(\vec{r},t) = (n_A - n_B)/(n_A +n_B)$. 
The quantity $\phi_{\textrm{av}}(z,t)$ is obtained by averaging 
$\phi(\vec{r},t)$ in the directions parallel to the surface, and then 
further averaging over 50 independent runs. These laterally-averaged 
profiles are analogous to depth profiles measured in experiments -- see 
Fig.~7 in Ref.~\cite{gnp97} or Fig.~4 in Ref.~\cite{wds03}. The 
enrichment profiles are shown for the case with field strength 
$\epsilon_a=2.0$ at times $t=280, 1400, 7000$. It is clear 
that a layer rich in $A$ forms at the surface immediately after 
the field is turned on. Due to the conservation of the order 
parameter, there must be a corresponding depletion layer which 
decays to $\phi_{\textrm{av}}\simeq0$ in the bulk. These profiles 
are in agreement with the experimental observations of Jones et al. 
\cite{jkrss89} on blends of deuterated and protonated polystyrene, 
and the experimental results of Mouritsen \cite{mouritsen} on 
biopolymer mixtures. Notice that similar profiles are seen for SDSD 
or surface-directed phase separation if the system is quenched to the 
metastable region of the phase diagram \cite{pb01}. The evolution dynamics 
in that case is analogous to the SE problem as long as droplets 
are not nucleated in the system.

We make the following observations about the enrichment profiles 
in Fig.~{\ref{fig:fig3}}. First, for the field strengths we consider, 
the surface is strongly enriched in $A$: $\phi_{\textrm{av}}(0,t)\simeq 1$ 
for $\epsilon_a=2.0, 3.0$. Thus, the linear theory is not applicable 
in the enrichment layer as this would require $\phi^3 \ll \phi$. 
Second, at late times (say $t=7000$ in Fig.~{\ref{fig:fig3}}), the 
depletion region stretches deep into the bulk. To avoid finite-size 
effects, we confine our simulation to time regimes where the 
thickness of the depletion region $\ll D=64$, the box size in the 
$z$-direction. As shown in Fig.~{\ref{fig:fig3}}, the profiles are 
fitted very well by the superposition of two exponential functions 
as in Eq.~(\ref{eq9:expfit}). Our simulations show that $B_-(t)$ and 
$\xi_-(t)$ rapidly saturate to their equilibrium values (in agreement 
with the predictions of linear theory \cite{bf91,pf93,fpn99}) and may 
be treated as static parameters. 

\begin{figure}[!htbp]
\begin{center}
\includegraphics*[width=0.95\textwidth]{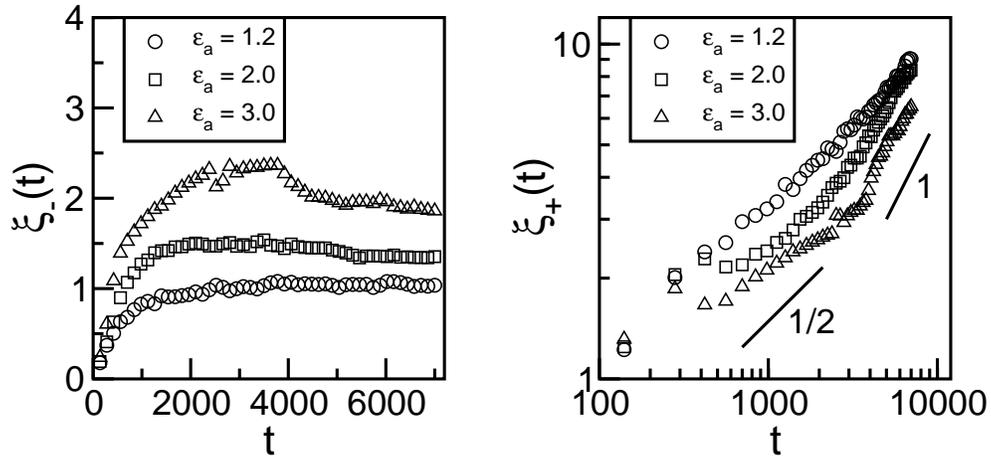}
\end{center}
\caption{Time-dependence of the fit parameters $\xi_-$ (left panel) 
and $\xi_+$ (right panel) of SE profiles. The surface field values 
are indicated in the legends. The length scale $\xi_-(t)$ rapidly 
saturates to its equilibrium value, whereas $\xi_+(t)$ shows a 
crossover from the diffusive regime ($\xi_+ \sim t^{1/2}$) to the 
hydrodynamic regime ($\xi_+ \sim t$) for higher surface field 
strengths ($\epsilon_a=2.0,3.0$). The lines of slope $1/2$ and $1$ 
are provided as a guide to the eye.}
\label{fig:fig4}
\end{figure}

In Fig.~{\ref{fig:fig4}}, we show the time-dependence of $\xi_-(t)$ 
and $\xi_+(t)$ for three surface field strengths, $\epsilon_a=1.2, 2.0, 3.0$. 
The saturation value of $\xi_-$ increases with the field strength. 
Further, we find that $\xi_+(t)$ grows with time as a power-law 
($\xi_+ \sim t^\theta$) 
but there is a clear crossover in the growth exponent. At early times 
($t\ll t_c$), we have $\theta \simeq 0.5$, in the conformity with the 
prediction of linear diffusive growth. However, there is much more rapid 
growth at late times ($t\gg t_c$) with $\theta \simeq 1.0$. To understand 
the crossover in Fig.~\ref{fig:fig4}, consider the dimensionless 
evolution equation for the order parameter in the presence of a velocity 
field $\vec{v}(\vec{r},t)$ \cite{bray94,pw09}:
\begin{eqnarray}
 \frac{\partial}{\partial t}\phi(\vec{r},t) &=& \nabla^2 \mu 
- \vec{v}\cdot \vec{\nabla}\phi , \label{eq15:phir} \\
 \mu &=& \phi + \phi^3 -\frac{1}{2} \nabla^2 \phi. \nonumber
\end{eqnarray}
For the SE problem, we set $\phi(\vec{r},t) \simeq \phi(z,t)$ and 
$v_z(\vec{r},t) \simeq v_z(z,t)$. Then 
\begin{equation}
 \frac{\partial}{\partial t}\phi(z,t) = \frac{\partial^2 \mu}{\partial z^2} 
- v_z \frac{\partial \phi}{\partial z} .
\label{eq16:phiz}
\end{equation}
We use the double-exponential form of $\phi(z,t)$ in Eq.~(\ref{eq9:expfit}) 
to estimate the various terms in Eq.~(\ref{eq16:phiz}) to leading order at 
$z \sim O(\xi_+)$, i.e., far from the surface. We have 
\begin{eqnarray}
 \frac{\partial \phi}{\partial t} & \sim & \frac{1}{\xi_+^2}\frac{d\xi_+}{dt}, \nonumber \\
 \frac{\partial^2 \mu}{\partial z^2} & \sim & \frac{1}{\xi_+^3}, \nonumber \\
 v_z \frac{\partial \phi}{\partial z} & \sim & \frac{v_z}{\xi_+^2}. \label{eq17:terms}
\end{eqnarray}
We have used the general relation $B_- \xi_- = B_+ \xi_+$ to obtain the
above expressions. In this case, the bulk is homogeneous and there is no structure 
formation in the density or velocity fields, i.e., $v_z \sim$ constant. 
At early times, the diffusive term in Eq.~(\ref{eq16:phiz}) dominates, 
yielding $\xi_+ \sim t^{1/2}$. At late times, the convective term in 
Eq.~(\ref{eq16:phiz}) is dominant, giving $\xi_+ \sim v_z t$. The precise 
dependence of the crossover on various physical parameters can be 
estimated by considering the dimensional version of Eq.~(\ref{eq15:phir}). 

The crossover in the growth exponent $\theta$ 
is reminiscent of phase-separation kinetics in fluid mixtures quenched 
below $T_c$ \cite{bray94,pw09}. In that case, the domains 
grow as $L(t) \sim t^x$ with the exponent crossing over from 1/3 
(diffusive regime) to 1 (viscous hydrodynamic regime) to 2/3 
(inertial hydrodynamic regime). Recently, we have observed the 
$1/3 \rightarrow 1$ crossover in the growth law for the wetting layer 
in MD studies of surface-directed phase separation \cite{jpd}. 
Our MD results in this paper show that convective transport accelerates 
the growth of the SE layer also. Further, the onset of 
the hydrodynamic regime is faster for stronger surface fields. This 
result has important experimental implications, and we urge experimentalists 
to undertake a detailed study of this problem.

\begin{figure}[!htbp]
\begin{center}
\includegraphics*[width=0.75\textwidth]{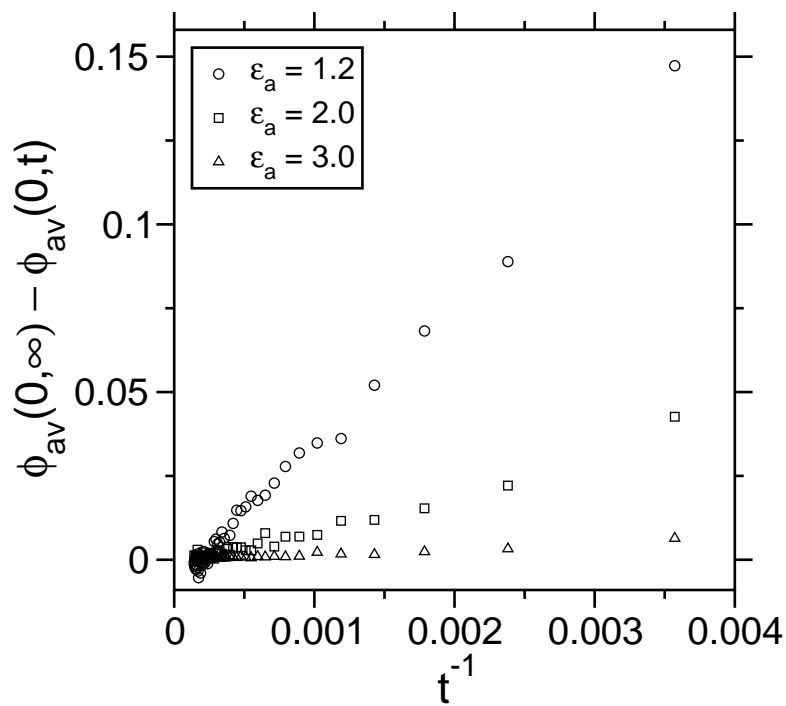}
\end{center}
\caption{Time-dependence of the surface value of the order parameter 
for the SE profiles. We plot 
$\phi_{\textrm{av}}(0,\infty)-\phi_{\textrm{av}}(0,t)$ vs. $t^{-1}$ 
for $\epsilon_a = 1.2, 2.0, 3.0$.}
\label{fig:fig5}
\end{figure}

We have also confirmed that $B_-(t)$ rapidly saturates to its equilibrium 
value, and that $B_+(t) \sim \xi_+(t)^{-1}$ (results not shown here). 
The profile parameters are consistent with the conservation constraint, 
$B_-\xi_- \simeq B_+\xi_+$. Figure~{\ref{fig:fig5}} shows the time-dependence 
of the surface value of the order parameter. We plot 
$\phi_{\textrm{av}}(0,\infty)-\phi_{\textrm{av}}(0,t)$ vs. $t^{-1}$ 
for $\epsilon_a=1.2, 2.0, 3.0$, demonstrating that $\phi_{\textrm{av}}(0,t)$ 
saturates linearly to its asymptotic value $\phi_{\textrm{av}}(0,\infty)$. 
Recall that $\phi(0,t) \simeq B_-(t) - B_+(t) \simeq a_1 - a_2/\xi_+$, 
and $\xi_+ \sim t$ in the late stages.

\begin{figure}[!htbp]
\begin{center}
\includegraphics*[width=0.75\textwidth]{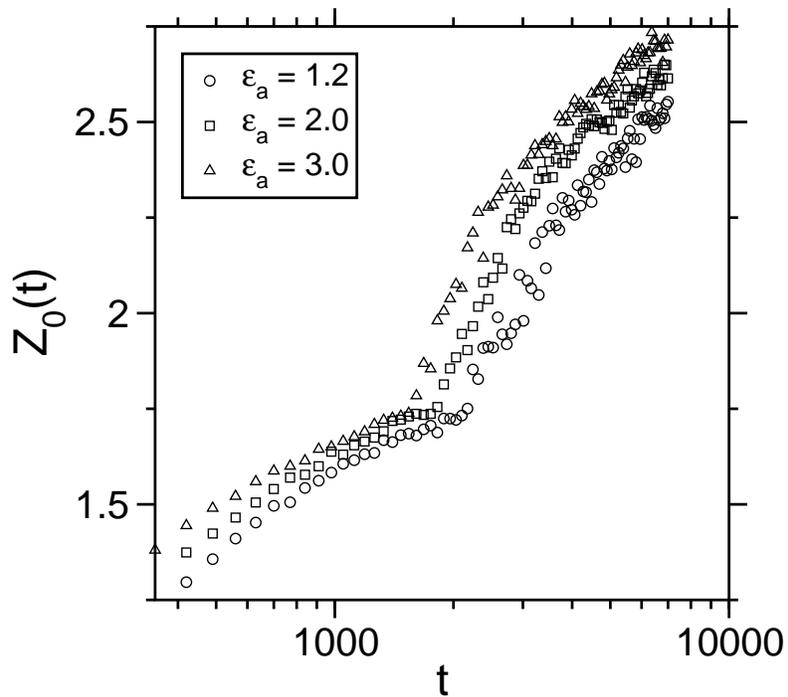}
\end{center}
\caption{Time-dependence of the first zero crossing of the SE profiles. 
We plot $Z_0(t)$ vs. $t$ (note the logarithmic scale of time) 
for $\epsilon_a=1.2,2.0,3.0$. This plot confirms the crossover in the 
growth exponent from the diffusive regime to the hydrodynamic regime.}
\label{fig:fig6}
\end{figure}

Finally, we focus on the time-dependence of the thickness of the enriched 
layer, which is also of considerable experimental interest. For the 
double-exponential profile in Eq.~(\ref{eq9:expfit}), the zero is located 
at [cf. Eq.~(\ref{eq13:1stzero})]
\begin{equation}
 Z_0(t)\simeq \xi_- \ln\left(\frac{B_-}{B_+}\right).
 \label{zero}
\end{equation}
Thus, we expect $Z_0(t)\sim\ln t$ in both the time-regimes of Fig.~{\ref{fig:fig4}}, 
but the slope should be steeper for $t>t_c$. This is precisely the 
behavior seen in Fig.~{\ref{fig:fig6}}, where we plot our MD results 
for $Z_0(t)$ vs. $t$ on a log-linear scale. This confirms that there 
is a crossover in the growth exponent from the diffusive regime 
($\theta \simeq 0.5$) to the hydrodynamic regime ($\theta \simeq 1.0$).

\section{Summary and Discussion}
\label{sec:sum}
Let us conclude this paper with a brief summary and discussion of our 
results. We are interested in the kinetics of surface enrichment (SE), 
which occurs when a miscible or metastable binary ($AB$) mixture is 
placed in contact with a surface having a preferential attraction for 
one of the components (say, $A$). A closely-related problem is that of 
surface-directed spinodal decomposition (SDSD), where an unstable 
homogeneous mixture is placed in contact with a wetting surface 
\cite{puri05}. The problems of SE and SDSD are of great scientific 
and technological importance. Experiments in this area have been 
performed on polymer blends, fluid mixtures, alloys, etc.

In this paper, we undertake comprehensive molecular dynamics (MD) 
simulations to study the kinetics of SE. The typical enrichment profile 
consists of an enriched surface layer, followed by an extended shallow 
depletion region. This profile propagates into the bulk with the 
passage of time. We are interested in understanding the role of 
hydrodynamic effects in driving the growth of the enrichment layer. 
In the context of phase-separation kinetics, we know that hydrodynamic 
transport drastically alters the intermediate and late stages of 
domain growth. At early times, the characteristic length scale 
of the SE profile grows diffusively with time ($\xi_+\sim t^{1/2}$), 
which is consistent with linear theory for the Cahn-Hilliard model. 
However, the growth exponent undergoes a crossover to a convective 
regime, and the late-stage dynamics is $\xi_+\sim t$. There is a 
corresponding crossover in the growth dynamics of the thickness of 
the enrichment layer $Z_0(t)$. The growth of $Z_0(t)$ is logarithmic 
in both regimes, but with a different slope.

The MD results presented here have significant implications for SE 
experiments, as many of these are performed on fluid mixtures. We 
hope that our MD results will provoke fresh experimental interest 
in this problem, and our theoretical results will be subjected to 
an experimental confirmation.

\vspace{0.5cm}
\noindent {\bf Acknowledgments}\\
\vspace{0.2cm}

PKJ acknowledges the University Grants Commission, India for financial 
support. SP is grateful to Kurt Binder and Harry Frisch for a fruitful 
collaboration on the problems discussed here.

\bibliography{draft}

\newpage

\newpage

\newpage

\newpage

\newpage

\newpage

\end{document}